# Monofluorinated Ether Electrolyte with Acetal Backbone for High-Performance Lithium Metal Batteries

**Authors**: Elizabeth Zhang[1,2, †], Yuelang Chen[1,3, †], Zhiao Yu[1,3], Yi Cui[2,4,5*], and Zhenan Bao[1*]

**Affiliations**:
[1]*Department of Chemical Engineering, Stanford University, Stanford, CA, USA.*
[2]*Department of Materials Science and Engineering, Stanford University, Stanford, CA, USA.*
[3]*Department of Chemistry, Stanford University, Stanford, CA, USA.*
[4]*Department of Materials Science and Engineering, Stanford University, Stanford, CA, USA.*
[5]*Stanford Institute for Materials and Energy Sciences, SLAC National Accelerator Laboratory, Menlo Park, CA, USA.*
[†]*These authors contributed equally: Elizabeth Zhang, Yuelang Chen*

*Corresponding authors: zbao@stanford.edu, yicui@stanford.edu

## Abstract

High degree of fluorination for ether electrolytes has resulted in improved cycling stability of lithium metal batteries (LMBs) due to stable SEI formation and good oxidative stability. However, the sluggish ion transport and environmental concerns of high fluorination degree drives the need to develop less fluorinated structures. Here, we introduce bis(2-fluoroethoxy)methane (F2DEM) which features monofluorination of the acetal backbone. High coulombic efficiency (CE) and stable long-term cycling in Li||Cu half cells can be achieved with F2DEM even under fast Li metal plating conditions. The performance of F2DEM is further compared with diethoxymethane (DEM) and 2-[2-(2,2-Difluoroethoxy)ethoxy]-1,1,1-Trifluoroethane (F5DEE). The structural similarity of DEM allows us to better probe the effects of monofluorination, while F5DEE is chosen as the one of the best performing LMB electrolytes for reference. The monofluorine substitution provides improved oxidation stability compared to non-fluorinated DEM, as demonstrated in the linear sweep voltammetry (LSV) and voltage holding experiments in Li||Pt and Li||Al cells. Higher ionic conductivity compared to F5DEE is also observed due to the decreased degree of fluorination. Furthermore, 1.75 M lithium bis(fluorosulfonyl)imide (LiFSI) / F2DEM displays significantly lower overpotential compared with the two reference electrolytes, which improves energy efficiency and enables its application in high-rate conditions. Comparative studies of F2DEM with DEM and F5DEE in anode-free (LiFePO$_4$) LFP pouch cells and high-loading LFP coin cells with 20 μm excess Li further show improved capacity retention of F2DEM electrolyte.

**Introduction**

Lithium metal has emerged as a highly promising battery anode material with its high theoretical specific capacity (3860 mAh g$^{-1}$) and low standard reduction potential (−3.04 V vs standard hydrogen electrode.[1–3] Despite their potential benefits, lithium metal batteries (LMB) still suffer from low coulombic efficiency (CE) and poor cycling stability.[4–6] Generally, if 1000 stable cycles with more than 90% capacity retention is desired, the averaged CE would have to be at least 99.99%.[7] One major factor that significantly impacts the CE is the formation of a stable SEI layer on the surface of the anode. The SEI layer is critical for preventing further reactions between the anode and electrolyte.[8,9] However, SEI is prone to cracking during cycling, which results in mossy Li growth, the formation of "dead Li," irreversible loss of lithium inventory, and excess SEI formation.[6]

Among the different strategies to modify the SEI formation and improve CE, rational electrolyte design is essential.[4,10–12] Some of the electrolyte engineering strategies that have been extensively investigated in recent years include high concentration electrolytes,[13] localized high concentration electrolytes,[14,15] additive design,[16] mixed solvent systems,[17,18] dual-salt-dual-solvent electrolytes,[19,20] and single-salt-single-solvent electrolytes.[12,21–23] Among the various electrolyte systems, fluoroethers have emerged as a promising class of solvent for lithium metal batteries.[12,22–28] Solvent fluorination not only modifies solvation structure for favorable SEI formation, but also improves oxidation stability with high-voltage cathodes. However, there are two major drawbacks. First, ionic conductivity in fluoroether electrolytes is significantly lower than the commercial carbonate electrolytes, which limits their operation to low charging rates. Second, the environmental concerns over the perfluorinated methyl and methylene carbons could limit their wide application in the battery industry.[29] Therefore, it is necessary to explore alternative ether derivatives with low degree of fluorination and high conductivity while maintaining favorable solvation structure for electrochemical stability.

To address the aforementioned issues, we designed and synthesized bis(2-fluoroethoxy)methane (F2DEM). The weakly solvating acetal backbone enables high Li CE. The monofluorine substitution on the end carbons improves oxidation stability, ionic conductivity, and further improves the Li CE. Compared with the two reference electrolytes, 1.7 M LiFSI / diethoxymethane (DEM) and 1.2 M LiFSI / 2-[2-(2,2-Difluoroethoxy)ethoxy]-1,1,1-Trifluoroethane (F5DEE), 1.75 M LiFSI / F2DEM showed excellent CE and stable long-term cycling in Li‖Cu half cells. A significantly lower overpotential was also observed with F2DEM, which improves the energy efficiency and enables its application in high-rate conditions. Additionally, F2DEM-based anode-free LFP pouch cells cycled under various charging and discharging conditions showed excellent performance, even under fast charge and slow discharge. Improved capacity retention was also observed with F2DEM in high-loading LFP coin cells with 20 μm excess Li, surpassing that of our previously reported high-performing F5DEE electrolyte.

## Molecular design

Ether solvents have demonstrated noticeable potential as electrolyte solvents in LMB due to their better stability with the Li anode as compared to carbonates.[21,30–32] However, their low oxidative stability poses challenges to applications in high-voltage batteries. To improve the cathode stability of ether electrolytes, incorporating electron-withdrawing fluorine atoms was found to deepen the HOMO energy level.[12,22,24–28] While a higher degree of fluorination (i.e. $CF_2$, $CF_3$) is expected to yield a higher oxidative stability, they are often accompanied by slower ion transport that leads to limited application in faster rate conditions.[22] Environmental concerns of highly fluorinated compounds also drive the need to develop monofluorinated alternatives.[29]

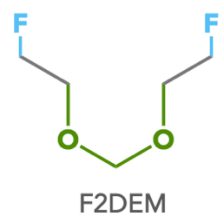

**Figure 1**. Molecular design of F2DEM that harvests the benefits of enhanced oxidative stability of monofluorination and weakened solvation of acetal backbone.

Monofluorinated ethers have been reported with ethylene glycol ether backbone, namely 1,2-bis(2-fluoroethoxy) ethane (FDEE).[28] However, when used as the single solvent with LiFSI, poor Li cycling stability was observed. To stabilize the CE of monofluorinated ethers, previous approach relied on a localized high concentration electrolyte (LHCE) design to reduce solvent participation in interfacial reactions and derive a more stable inorganic-rich SEI.[28] However, the LHCE approach not only requires highly fluorinated compounds with great environmental concerns, but the addition of highly fluorinated diluents can also hinder the ion transport and lower the overall conductivity of the electrolyte. These drawbacks motivate us to test the feasibility of designing a single-solvent system that achieves excellent Li cycling performance even with low degree of fluorination.

Our recent work suggested that the acetal structure can effectively weaken the solvation power. We hypothesize that this will enable more stable cycling even in monofluorinated systems, without having to increase the concentration of the electrolyte or introducing highly fluorinated diluent. Based on these design principles, F2DEM is synthesized with monofluorination of the acetal backbone (Figure 1), which ensures stability at both the anode and cathode side, in addition to maintaining good ionic conductivity.

## Electrolyte characterization

Adequate ion transport is essential to a high-performing LMB system, and Li salt concentrations can affect the ionic conductivity.[33] Therefore, it is necessary to first determine the optimal salt concentration before proceeding with cell cycling and characterizations. Electrolyte solutions were obtained by mixing LiFSI with F2DEM at 1.2 to 3 mol LiFSI / L solvent and ionic conductivities were measured without and with a separator. The setup without a separator allows us to measure the intrinsic conductivity of the electrolytes, while conductivity with separator better mimics the condition in realistic cells. For the setup with separator, SS||SS coin cells were assembled with Celgard 2325 separator swelled by different concentrations F2DEM. To measure the conductivity without separator, Swaglok cells were used. The ionic conductivity with separator peaked at around 2 mol LiFSI / L F2DEM, which correlates to 1.75 M LiFSI / F2DEM (Figure 2a). In the separator-free Swaglok setup, 1.75 M remained the optimal concentration for high ionic conductivity (Figure 2b). The transport numbers of different concentrations were also measured for F2DEM. Higher transport number is generally desired as a higher fraction of the current is carried by Li$^+$. As shown in Figure 2c, the measured values for all concentrations (> 0.4) are comparable to the transport number commonly reported in ether electrolytes.[12,22,27]

Overpotential is another critical parameter in LMB as it is directly related to the high-rate performance and energy efficiency of the system. To assess the overpotential of F2DEM, Li||Li symmetric cells were made with thick Li (750 μm) on both sides. The cells were cycled under various current densities. With a capacity of 3 mAh cm$^{-2}$ for each cycle, the current density was gradually increased from 1 mA cm$^{-2}$ to 8 mA cm$^{-2}$, where the cells were cycled 10 times under each current density. As shown on Figure 2d, the overpotential of 1.75 M LiFSI / F2DEM is roughly 50% less than that of 1.2 M LiFSI/ F5DEE under all current densities. Under a current density of 1 mA cm$^{-2}$, an overpotential of around 55 mV was observed for F5DEE, similar to our previous report.[22] The overpotential of F2DEM, however, is only around 30 mV. This significant drop in overpotential suggests a much higher energy efficiency of F2DEM over F5DEE. This set up also simultaneously assesses the fast-charging capability of the F2DEM electrolyte. Zooming in on the voltage profile under 6 mA cm$^{-2}$, soft short was observed for both F2DEM and F5DEE, indicating that the system may be unstable under fast rates over 6 mA cm$^{-2}$.

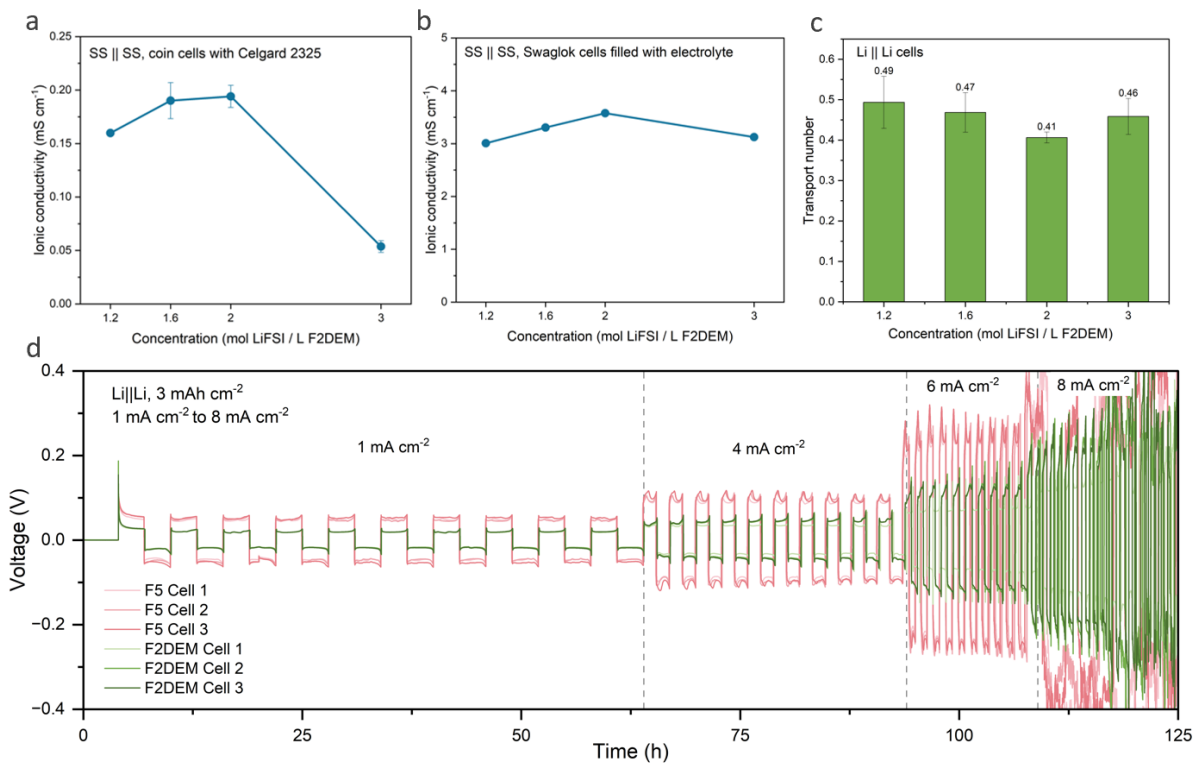

**Figure 2.** The transport properties of F2DEM. (a) Ionic conductivity of F2DEM measured with Celgard 2325 separator in SS || SS cells across four different concentrations. The ionic conductivity values are averaged among three repeated cells and the standard deviations are shown on the plot. (b) The intrinsic ionic conductivity without a separator measured with Swaglok cells across four different concentrations. (c) Transport numbers measured in Li||Li symmetric cells for the four different concentrations. The transport numbers are averaged among three repeated cells and the standard deviations are shown on the plot. (d) Li||Li symmetric cells cycled at increasing current densities (1 mA cm$^{-2}$ to 8 mA cm$^{-2}$) and 3 mAh cm$^{-2}$ capacity. The cells are cycled 10 times under each current density.

**Electrochemical stability**

Next, we investigate the performance of 1.75 M LiFSI / F2DEM in Li||Cu half cells as compared to 1.2 M LiFSI/ F5DEE and 1.7 M LiFSI / DEM. The average coulombic efficiency (CE) was first evaluated by Aurbach method in Li||Cu half cells[34,35] (Figure 3a). Based on this standard protocol, 5 mAh cm$^{-2}$ of Li was first deposited onto the Cu foil as Li reservoir. This was followed by 10 subsequent cycles of plating and stripping at 0.5 mA cm$^{-2}$ for 1 mAh cm$^{-2}$. Finally, all deposited Li was stripped from Cu, and the total capacity recovered was divided by the amount deposited to obtain the CE. The average CE of F2DEM was measured as 99.53 ± 0.03% for four cells, which is higher than the 99.46 ± 0.01% (3 cells) previously observed with F5DEE[22] and 99.19 ± 0.07% (4 cells) with DEM. The high CE of F2DEM is indicative of the excellent stability at the Li anode.

Furthermore, we corroborate these observations through long-term cycling of Li‖Cu half cells. To probe the long-term cycling stability, the Cu surface was first conditioned by cycling between 0 and 1 V at 0.2 mA cm$^{-2}$ for 10 cycles. For long-term cycling, 1 mAh cm$^{-2}$ of Li was plated onto Cu at 0.5 mA cm$^{-2}$, and stripped to 1 V at a rate of 0.5 mA cm$^{-2}$. Zooming in on the first 50 cycles (Figure 3b), fast activation (number of cycles required to reach 99%) was observed for F2DEM, indicating that less capacity was lost during the initial cycles toward the establishment of a stable SEI. This is highly desirable for anode-free cells with limited Li supply. Excluding the activation cycles, the average CE of F2DEM in the first 5 to 50 cycles (99.44%) was also higher than that of F5DEE and DEM (both at 99.29%). Similar experiments were carried out under a faster charge and slower discharge condition (1 mA cm$^{-2}$ plating, 0.4 mA cm$^{-2}$ stripping) with a higher capacity (2 mAh cm$^{-2}$) to further evaluate the fast-charging capabilities of F2DEM. Under this harsher condition, a more distinct difference can be observed among the three different electrolytes. F2DEM showed a superior cycling stability with a higher CE over 250 cycles compared with F5DEE and DEM (Figure 3c). Comparing to the long-term cycling at 0.5 mA cm$^{-2}$, this fast plating and slow stripping condition requires more number of cycles to reach stable cycling at 99% for all three electrolytes. However, a more distinct difference can be observed. While F2DEM only took around 12 cycles to reach 99% CE, F5DEE took over 25 cycles to establish a stable SEI and DEM fails to achieve a stable cycling over 99%. These results suggest that F2DEM is more stable against the Li anode under higher rate conditions.

Considering the modified solvation environment and addition of electro-withdrawing fluorine atoms in F2DEM molecules, we would expect an improved oxidative stability of the 1.75 M F2DEM. To corroborate the oxidative stability enhancement of F2DEM upon fluorination, we carried out linear sweep voltammetry (LSV) on the three different electrolytes. To best mimic the realistic full cell cycling environment, Al was first chosen as the working electrode. Sweeping up to 7 V at 1 mV s$^{-1}$, the resulting leakage current would allow us to evaluate the corrosion of Al current collector.[36,37] With this set up, we observed a leakage current onset at around 6 V for both F2DEM and DEM (Figure 3d). The delayed voltage onset around 6.5 V for F5DEE is likely because of its increased degree of fluorination, which results in a higher oxidative stability.

In addition to LSV, the electrode/electrolyte stability is further assessed through voltage-holding experiments in the Li‖Al cell setup. In this case, the voltage was slowly increased from an open circuit to 4.4 V at a scan rate of 1 mV s$^{-1}$, then the cells were held at 4.4 V for an extended period of time, and the corresponding leakage current was monitored. In this case, no significant leakage current should be observed while holding at 4.4 V if a stable passivation layer was established on the Al surface.[21] As shown in Figure 3f, we did not observe a significant leakage current for all three electrolytes. Overall, the passivation of Al is relatively stable at 4.4 V for the tested systems.

LSV experiments were also conducted with Pt working electrode. Unlike the Al cathode current collector, no passivation reactions occured on the Pt surface. With the absence of a passivating layer, the onset voltage is expected to decrease for all three electrolytes. As shown in Figure 3e, the leakage current was observed at an onset voltage of around 4 V for F2DEM, which is higher than the 3 V observed with DEM. With a higher degree of fluorination in F5DEE, it expectedly shows a higher onset voltage at around 4.5 V. This trend corroborates our hypothesis that increasing the degree of fluorination enhances the oxidative stability and agrees well with the previous research.[22]

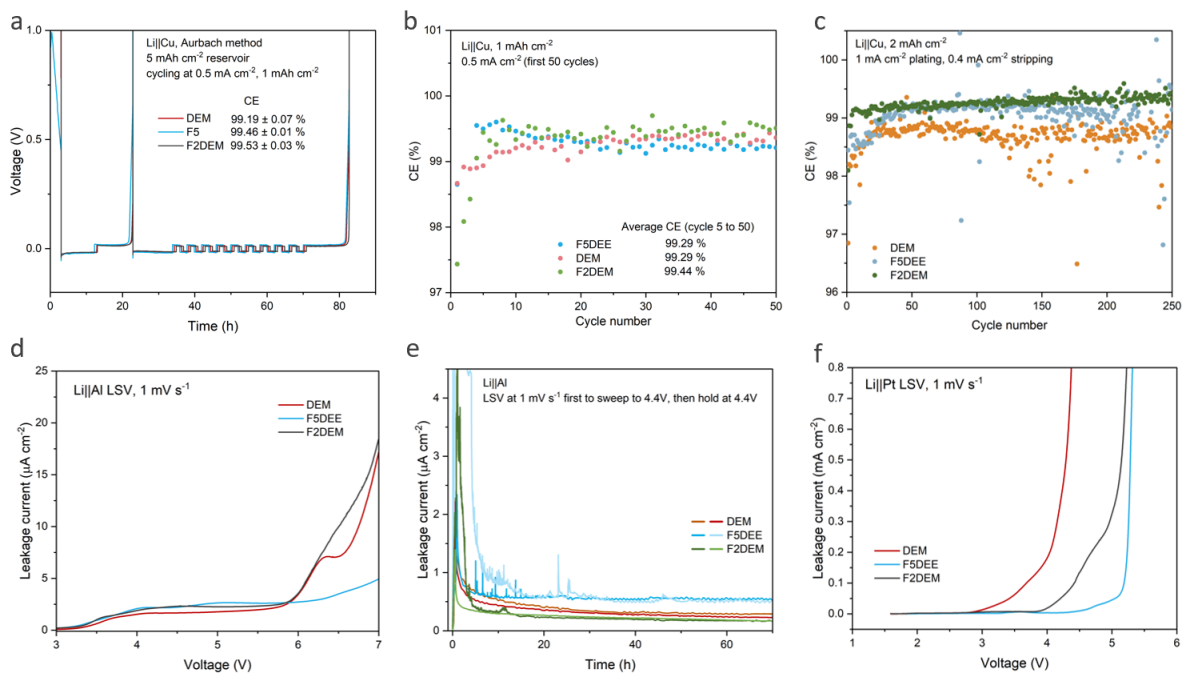

**Figure 3.** Electrochemical stability of 1.75 M LiFSI / F2DEM compared with 1.7 M LiFSI / DEM and 1.2 M LiFSI / F5DEE. (a) The CE measurements of the three electrolytes based on the modified Aurbach method.[35] The CE was averaged among four cells for F2DEM and DEM and the standard deviations are shown on the plot. For F5DEE, the CE was averaged between two cells and the standard deviation is shown on the plot. (b) Li||Cu CE of all three electrolytes. The cells were cycled at 0.5 mA cm$^{-2}$ for 1 mAh cm$^{-2}$ and stripped to 1 V at 0.5 mA cm$^{-2}$. Note that prior to cycling, the copper surface is pre-conditioned by cycling between 0 and 1 V at 0.2 mA cm$^{-2}$ for 10 cycles. The figure shows the CE of the first 50 cycles. The average CE values were calculated based on cycle 5 to 50, excluding the activation cycles. Four repeated cells for F2DEM and DEM, and two repeated cells for F5DEE are made, showing the consistent trend. (c) Li||Cu CE over 250 cycles under a fast plating (1 mA cm$^{-2}$) and slow stripping (0.4 mA cm$^{-2}$) condition. Four repeated cells for F2DEM and DEM, and two repeated cells for F5DEE are made, showing the consistent trend. (d) The linear sweep voltammetry (LSV) in Li||Al cells. The leakage current of the three electrolytes were measured by sweeping up to 7 V at 1 mV s$^{-1}$. (e) Leakage current measured in Li||Al cells. LSV was first applied to the cells to sweep from open circuit voltage to 4.4 V at 1 mV s$^{-1}$. The cells were then held at 4.4 V for over 60 hours. Two repeated cells are tested for each electrolyte and similar trend is observed. (f) LSV in Li‖Pt cells showing improved oxidative stability comparing to DEM. The leakage currents of the three electrolytes were measured by sweeping up to 7 V at 1 mV s$^{-1}$.

**Pouch cell performance**

The performance of 1.75 M LiFSI / F2DEM was further assessed in Cu‖LFP anode-free pouch cells, with a voltage range from 2.5 V to 3.65 V. Various charging and discharging rates were implemented to fully assess the pouch cell performance of F2DEM (1C = 200 mA or 2 mA cm$^{-2}$). Note that the nominal capacity at a C/3 charge rate was 210 mAh, or 2.1 mAh cm$^{-2}$, and the electrolyte loading was 0.5 mL. We first studied the performance of 1.75 M LiFSI / F2DEM under C/2 charge and 2C discharge rate, since slow charge and fast discharge conditions were generally implemented to enhance the Li morphology. This was compared with the 1.2 M LiFSI / F4DEE and 1.2 M LiFSI / F5DEE electrolytes. Among the three electrolytes, F2DEM showed a higher capacity utilization and a slower capacity loss (Figure 4a–b). This improved capacity retention of F2DEM can be partially attributed to its low overpotential and higher ionic conductivity, which is consistent with our previous observation in Li‖Li symmetric cells (Figure 2d). With C/2 charge and a slower C/5 discharge rate, significant improvement in discharge capacity and CE was observed in 1.75 M LiFSI / F2DEM compared to 1.2 M LiFSI / F5DEE and F4DEE (Figure 4c–d). This is the more demanding condition since it has been reported that slower discharge than charge can lead to higher surface area Li morphology.[38] The improvement seen in F2DEM under a slower discharge rate potentially indicates that F2DEM can facilitate the formation of a more stable SEI. Under symmetric C/2 charge rate and C/2 discharge rate, 1.75 M LiFSI / F2DEM yielded similar cycling performance as the two reference electrolytes (Figure 4e–f). This is likely because the advantage of low overpotential cannot be clearly observed under slow conditions. Overall, 1.75 M LiFSI / F2DEM showed excellent performance in the anode-free LFP pouch cells under all tested rates, and particularly under fast charge and slow discharge condition where a more stable SEI is required to enhance the cycling stability.

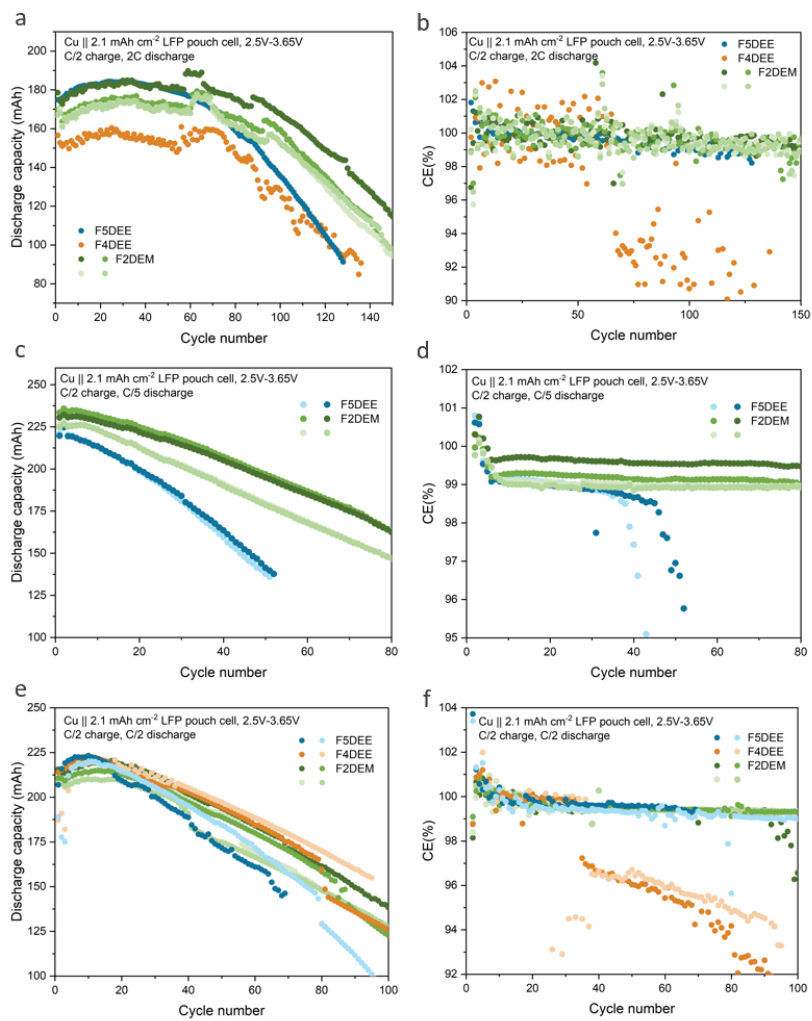

**Figure 4.** Performance of Cu||LFP pouch cell cycling between 2.5 V and 3.65 V. The nominal capacity at C/3 is 210 mAh, or 2.1 mAh cm$^{-2}$. The electrolyte loading is 0.5 mL. The 1.75 M LiFSI / F2DEM is compared to 1.2 M LiFSI / F5DEE and F4DEE under various charging and discharging conditions. (a, b) C/2 charge and 2C discharge capacity (1C = 200 mA or 2 mA cm$^{-2}$) and CE profile over 80 cycles. Four repeated cells for F2DEM are shown. (c, d) C/2 charge and C/5 discharge capacity and CE profile over 150 cycles. Two repeated cells for F5DEE and four repeated cells for F2DEM are shown. (e, f) C/2 charge and C/2 discharge capacity and CE profile over 100 cycles. Two repeated cells for F5DEE and F4DEE, and four repeated cells for F2DEM are shown. The data for F4DEE and F5DEE are taken from Ref. 22.

**Coin cell performance**

The cycling performance of 1.75 M LiFSI / F2DEM was also evaluated in Li||LFP coin cells with 20-μm-thick Li anode and high-loading 3.5 mAh cm$^{-2}$ LFP cathode. The additional lithium source

will allow us to probe the long-term cycling performance of Li||LFP cells. Various charge and discharge current densities (0.75 mA cm$^{-2}$ charge and 1.5 mA cm$^{-2}$ discharge, 1.5 mA cm$^{-2}$ charge and 3 mA cm$^{-2}$ discharge, 0.4 mA cm$^{-2}$ charge and 2 mA cm$^{-2}$ discharge, and 1mA cm$^{-2}$ charge and 2 mA cm$^{-2}$ discharge) were applied with 3.8 V cutoff. With 0.75 mA cm$^{-2}$ charge and 1.5 mA cm$^{-2}$ discharge, a higher capacity retention was observed for F2DEM than F5DEE and DEM (Figure 5a–b). From the charge and discharge curves of all three electrolytes (Figure 6a–c), the voltage plateau increased over cycling, indicating an increase in overpotential. However, since voltage divergence can still be observed after 150 cycles, overpotential increase should not be the only failure mechanism. The lithium inventory loss may also contribute to the capacity loss. Therefore, the improved capacity retention of F2DEM is likely due to its low and stable overpotential, in combination with its formation of a more stable SEI to reduce lithium inventory loss, prolonging the cycle life of F2DEM-based cells. Further increasing the current densities to 1.5 mA cm$^{-2}$ charge and 3 mA cm$^{-2}$ discharge, F2DEM can retain 80% of the initial capacity over 125 cycles while the capacity of F5DEE dropped below 80% retention after 100 cycles (Figure 5c–d). The charge and discharge curves (Figure 6d–e) showed a similar trend as the 0.75 mA cm$^{-2}$ charge and 1.5 mA cm$^{-2}$ discharge condition, where F2DEM showed superior capacity retention with a lower overpotential increase and a less lithium inventory loss. DEM was not included in this comparison due to its poor performance even under slower charging conditions. The performance of F2DEM is also assessed in 0.4 mA cm$^{-2}$ charge and 2 mA cm$^{-2}$ discharge, as well as 1 mA cm$^{-2}$ charge and 2 mA cm$^{-2}$ discharge conditions. Under a relatively fast discharge rate of 2 mA cm$^{-2}$, F2DEM can retain 80% of its first cycle capacity over 350 cycles with 0.4 mA cm$^{-2}$ charge rate (Figure 5e–f). When the charge rate was increased to 1 mA cm$^{-2}$, the cells can still achieve stable CE and 80% capacity retention over 200 cycles (Figure 5g–h). These observations corroborate F2DEM's excellent stability even under high-rate conditions.

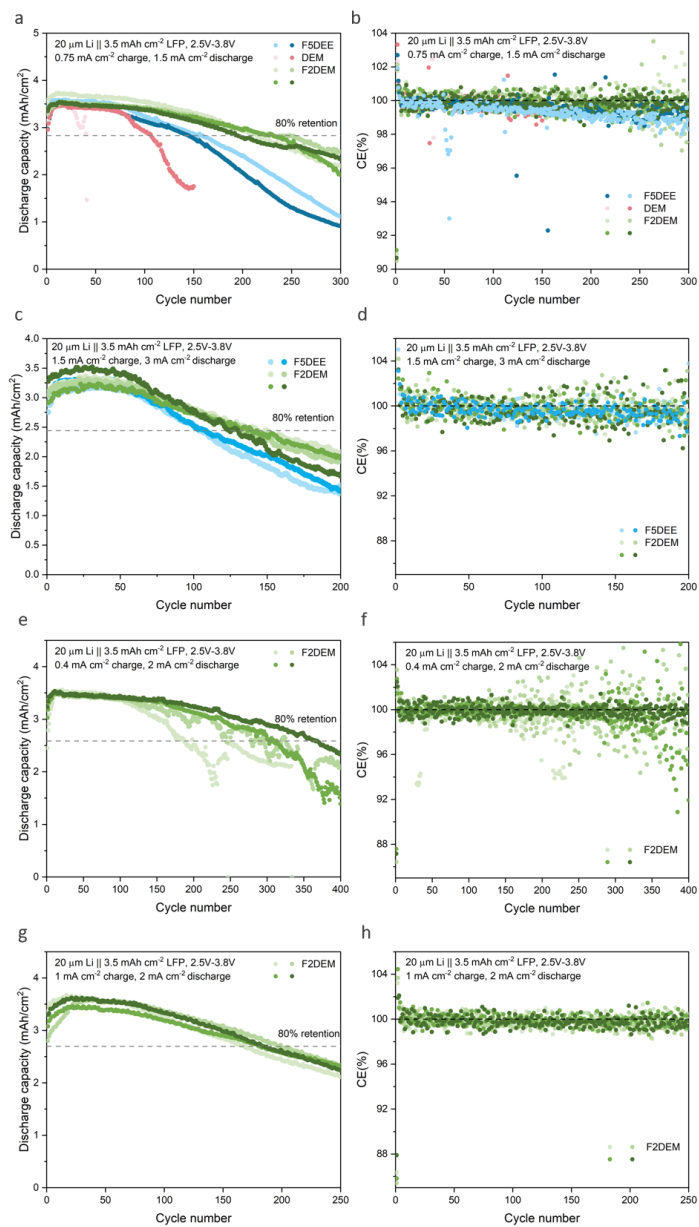

**Figure 5.** Performance of Li||LFP cells with 20-μm Li anode and high-loading 3.5 mAh cm$^{-2}$ LFP cathode for 1.75 M LiFSI / F2DEM, 1.2 M LiFSI / F5DEE, and 1.7 M LiFSI / DEM. The cells are cycled between 2.5 V to 3.8 V. The 80% capacity retention is defined by setting the cell with highest first cycle discharge capacity as the 100% capacity reference. (a, b) Discharge capacity and CE profile of cells cycled under 0.75 mA cm$^{-2}$ charge and 1.5 mA cm$^{-2}$ discharge (with comparison to F5DEE and DEM). Two repeated cells for F5DEE and DEM, and four repeated cells for F2DEM are shown. (c, d) Discharge capacity and CE profile of cells cycled under 1.5 mA cm$^{-2}$ charge and 3 mA cm$^{-2}$ discharge (with comparison to F5DEE). Two repeated cells for F5DEE and four repeated cells for F2DEM are shown. (e, f) Discharge capacity and CE profile of F2DEM cells cycled under 0.4 mA cm$^{-2}$ charge and 2 mA cm$^{-2}$ discharge. Four repeated cells for F2DEM

are shown. (g, h) Discharge capacity and CE profile of F2DEM cells cycled under 1 mA cm$^{-2}$ charge and 2 mA cm$^{-2}$ discharge. Four repeated cells for F2DEM are shown.

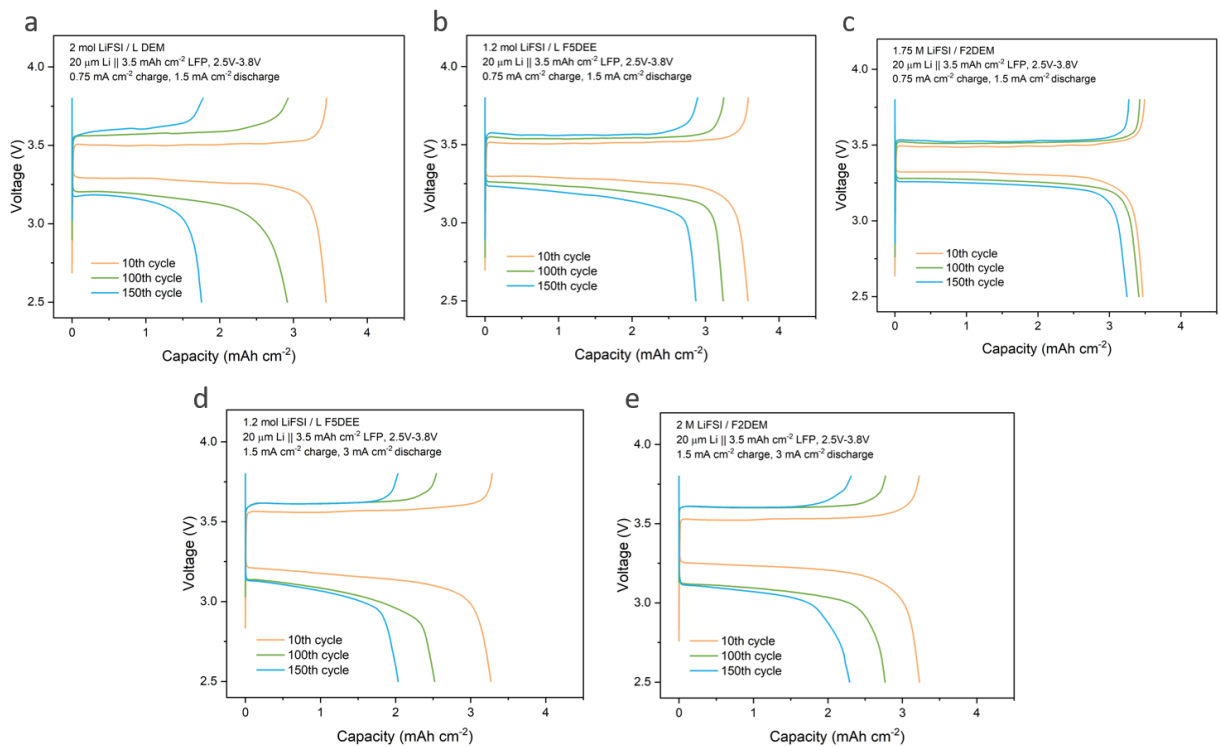

**Figure 6.** Charge / discharge curves of Li||LFP cells with 20-μm Li anode and high-loading 3.5 mAh cm$^{-2}$ LFP cathode. All cells are cycled between 2.5 V to 3.8 V. (a, b, c) Charge / discharge curves of cycle 10, 100, and 150 for 1.7 M LiFSI / DEM, 1.2 M LiFSI / F5DEE, and 1.75 M LiFSI / F2DEM, respectively. The cells were cycled under 0.75 mA cm$^{-2}$ charge and 1.5 mA cm$^{-2}$ discharge. The same trend is observed for the two repeated cells for F5DEE and DEM, and four repeated cells for F2DEM. The plot is based on the cell with the best capacity retention for all three electrolytes. (d, e) Charge / discharge curves of cycle 10, 100, and 150 for 1.2 M LiFSI / F5DEE and 1.75 M LiFSI / F2DEM, respectively. The cells were cycled under 1.5 mA cm$^{-2}$ charge and 3 mA cm$^{-2}$ discharge. The same trend is observed for the two repeated cells for F5DEE and four repeated cells for F2DEM. The plot is based on the cell with the best capacity retention for the two electrolytes.

**Conclusion**

In summary, we found that a monofluorinated ether electrolyte (F2DEM) is effective in improving the coulombic efficiency (CE) and cycling stability of LMBs. The weakly solvating acetal backbone promotes a stable solid electrolyte interface (SEI) formation and lower overpotential, while the monofluorine substitution on the end carbons further improves oxidation stability, ionic conductivity, and Li passivation. These modifications enabled high Li CEs and stable long-term cycling in Li||Cu half cells, even under fast plating and slow stripping conditions. Compared to

reference electrolytes (1.2 M LiFSI / F5DEE and 1.7 M LiFSI / DEM), 1.75 M LiFSI / F2DEM exhibited good ionic conductivity, high transport number, and significantly lower overpotential. Furthermore, comparative studies in anode-free LFP pouch cells and high-loading LFP coin cells with 20 μm excess Li demonstrated that F2DEM-based systems with improved capacity retention compared to the reference electrolytes under various charging and discharging rates. These results indicate the potential of monofluorinated DEM for enhancing the performance of LMBs.

**General Materials**

F5DEE was provided by Feon Energy. 2-Fluoroethanol was purchased from Matrix Scientific. 2,2-Difluoroethanol was purchased from SynQuest. NaOH, tetraglyme, dibromomethane, DEM and other general reagents were purchased from Sigma Aldrich or Fisher. The separator Celgard 2325 (25-μm thick, polypropylene/polyethylene/polypropylene) was purchased from Celgard. Thick Li foil (roughly 750-μm thick) and Cu current collector (25-μm thick) were purchased from Alfa Aesar. Thin Li foils (roughly 50- and 20-μm thick, supported on Cu substrate) were purchased from China Energy Lithium. Commercial LFP cathode sheets were purchased from Targray. Industrial dry Cu||NMC532 and Cu||LFP pouch cells were purchased from Li-Fun Technology. Other battery materials, such as 2032-type coin-cell cases, springs and spacers, were all purchased from MTI. All materials were used as received.

**Electrolyte synthesis**

To a 500 mL round flask was added 64 g of 2-fluoroethanol, 85 g of dibromomethane, 43 g of NaOH and 200 mL tetraglyme. The suspension was stirred at room temperature for 2 h under air and then heated to 40 °C to stir overnight. The suspension turned brownish with yellow fine powder. The suspension was further heated to 70 °C to stir overnight. The suspension was directly distilled under vacuum (vapor temperature ~70-75 °C at ~1 kPa) to obtain colorless liquid as the product. The crude product was distilled under vacuum for four times to ensure purity.

**Electrochemical Measurements**

All battery components used in this work were commercially available and all electrochemical tests were carried out using 2032-type coin cells. The cells were fabricated in an argon-filled glovebox, and one layer of Celgard 2325 was used as a separator for all batteries. Thick Li foil (750 μm) with a diameter of 7/16 in. was used for cell assembly unless otherwise specified, and 40 μL of electrolyte was injected in all cells with Cu||LFP as an exception. Both the ionic conductivity and transport number of the electrolytes were measured with Biologic VSP system. Ionic conductivity was derived from bulk impedance in symmetric cells with two stainless steel electrodes and electrolyte soaked separator. Li+ transport number was obtained by a Li||Li symmetric cell under a polarization potential of 10 mV. Li||Cu, Li||Li, Li||LFP cells were tested on

Land battery testing station, and Cu‖LFP pouch cells were tested on Arbin. The Li‖Li cells were cycled under different charge and discharge current densities (1 mA cm$^{-2}$, 4 mA cm$^{-2}$, 6 mA cm$^{-2}$, 8 mA cm$^{-2}$). The CEs of the electrolytes were measured based on a modified Aurbach method[35] in Li‖Cu cells, where 5 mAh cm$^{-2}$ of Li was first deposited onto the Cu foil as Li reservoir. This was followed by 10 subsequent cycles of plating and stripping at 0.5 mA cm$^{-2}$ for 1 mAh cm$^{-2}$. Finally, all deposited Li was stripped from Cu, and the total capacity recovered was divided by the amount deposited to obtain the CE. The long-term Li‖Cu cycling were carried out by first conditioning the Cu surface through cycling between 0 to 1V at 0.2 mA cm$^{-2}$ for 10 cycles. Upon preconditioning the Cu, 1 mAh cm$^{-2}$ of Li was plated on the Cu at 0.5 mA cm$^{-2}$ and stripped to 1V at the same rate. A faster plating and slower stripping condition was also used to further evaluate the cells, where 2 mAh cm$^{-2}$ of Li was plated on Cu at 1 mA cm$^{-2}$ and stripped to 1V at 0.4 mA cm$^{-2}$. The linear sweep voltammetry (LSV) was also carried out in Li‖Al and Li‖Pt cells using Biologic VSP300. In both setups, the voltage was swept from open-circuit value to 7V vs Li$^+$/Li at 1 mV s$^{-1}$. The leakage current were evaluated by dividing the measured value by the electrode area of 2.11 cm$^{-2}$. Al corrosion tests were carried out in Li‖Al cells, where the voltage is held at 4.4V for over 60 hours. The Cu‖LFP pouch cells were tested with 0.5 mL electrolyte injected into the purchased cells. The pouch cells were clamped in woodworking vises to ensure an estimated pressure of 1000 kPa and cycled under various charge and discharge conditions between 2.5V and 3.65V. The Li‖LFP coin cells were assembled with 20μm Li and 3.5 mAh cm$^{-2}$ LFP. The cells were cycled between 2.5V and 3.8V under different rates after a formation cycle at 0.4 mA cm$^{-2}$ charge and 1.5 mA cm$^{-2}$/3 mA cm$^{-2}$ discharge currents. The Al-clad cathode cases were also used for Li‖LFP coin cells, any defects in the Al cladding are expected to be minimized with Al foil inserted in the cathode case.

**Author information**
**Corresponding Authors**
Zhenan Bao − Department of Chemical Engineering, Stanford University, Stanford, California 94305, United States; orcid.org/0000-0002-0972-1715; Email: zbao@ stanford.edu

Yi Cui − Department of Materials Science and Engineering, Stanford University, Stanford, California 94305, United States; Stanford Institute for Materials and Energy Sciences, SLAC National Accelerator Laboratory, Menlo Park, California 94305, United States; orcid.org/0000-00026103-6352; Email: yicui@stanford.edu

**Authors**
Elizabeth Zhang – Department of Chemical Engineering, Stanford University, Stanford, California 94305, United States; Department of Materials Science and Engineering, Stanford University, Stanford, California 94305, United States; orcid.org/0000-0002-1117-8635

Yuelang Chen − Department of Chemical Engineering, Stanford University, Stanford, California 94305, United States; Department of Chemistry, Stanford University, Stanford, California 94305, United States; orcid.org/ 0000-0002-5249-0596

Zhiao Yu − Department of Chemical Engineering, Stanford University, Stanford, California 94305, United States; Department of Chemistry, Stanford University, Stanford, California 94305, United States; orcid.org/0000-00018746-1640

**Author Contributions**
E.Z. and Y.Chen. contributed equally to this work.

**Acknowledgment**
The work was supported by the Assistant Secretary for Energy Efficiency and Renewable Energy, Office of Vehicle Technologies of the U.S. Department of Energy under the Battery 500 Consortium program. E. Zhang acknowledges the support from National Science Foundation Graduate Research Fellowships Program (NSF GRFP). Y.Chen acknowledges the support from Chevron Fellowship. Part of this work was performed at the Stanford Nano Shared Facilities (SNSF), supported by the National Science Foundation under Award ECCS-2026822. We thank Feon Energy for providing F5DEE.

**Conflict of Interest**
This work has been filed as PCT application US23/21234.